# Novel STT/SHE MTJ Compact Model Compatible with NGSPICE

Jagadish Rajpoot, *Student Member*, *IEEE,* Ravneet Paul, and Shivam Verma, *Member*, *IEEE*
Department of Electronics Engineering, Indian Institute of Technology (BHU), Varanasi, 221005, India

**Abstract-** Ensuring high performance, while meeting the power budget is a challenging task as the world is moving towards next-generation computing. Researchers and designers are in search of new solutions for efficient computation. Spintronics devices have been viewed as a promising way to deal with the escalating difficulties of CMOS downscaling, explicitly, the Magnetic Tunnel Junction (MTJ) devices have been the focal point of investigation. They possess some essential features from the aforementioned perspective such as nonvolatility, low power, and scalability. In light of the significance of MTJ devices in next-generation computing, this paper presents a physics-based STT/SHE MTJ model for hybrid MTJ/CMOS circuit simulation, that accurately emulates the device physics and stochastic thermal noise behavior of the MTJ. It is vital to have an MTJ compact model which is compatible with the open-source NGSPICE simulation framework since previously developed models are reliant on commercial EDA tools. In addition, for developing hybrid circuits with random process fluctuations, a simulator-independent Monte-Carlo simulation capability has been incorporated Finally, the STT/SHE-MTJ model is demonstrated using PCSA read/write operation and the implementation of neuron MTJ.

## I. INTRODUCTION

Magnetic tunnel junction (MTJ) devices have garnered considerable attention in recent years because of their appealing characteristics, such as nonvolatility, scalability, and low power consumption [1-3]. These devices have been integrated with CMOS technology to develop spin-transfer torque random access memories (STT-MRAM) and spin-Hall effect magnetoresistive random access memory (SHE-MRAM). MTJ devices have proved their relevance by offering the non-volatility over static and dynamic random-access memory technologies in recent years. Furthermore, researchers are also working towards new nonvolatile logic circuits such as in-memory [4] and neuromorphic [5] computing. Hence, the MTJ devices can be considered quite prominent for future logic and non-volatile memory applications.

CMOS-based circuits are commonly used to select, program, and read out MTJ devices. Hence, there is need to efficiently simulate hybrid MTJ/CMOS devices, circuits, and systems to evaluate their performance before fabrication. Therefore, there is a need to develop a SPICE-compatible MTJ model which meticulously captures both magnetic and electrical characteristics of MTJ.

Several MTJ models with various features and implementation methods have been introduced such as micro-magnetic, behavioral, commercialized-tool-based, and macro model. Micro-magnetic model [6] accurately captures magnetization dynamics of a MTJ however, their use is limited for circuit design due to the requirement of high computation, Compact MTJ model uses a hardware description language, such as Verilog-A [7], to describe the analog behavior of an MTJ. These Verilog-A MTJ models are simulated using commercial software such as HSPICE and Cadence. Commercial tool-based model [8] such as Sentaurus device provide reasonable simulation accuracy however complex hybrid MTJ/CMOS circuit implementation is difficult. Further commercialized tools usage during pandemic becomes difficult due to inaccessibility to labs/research facilities whereas open-source tools are easily accessible. MTJ models that have been proposed in the past, all of them rely on commercial EDA tool access, which is costly and might be a barrier to research in this area. In this paper, an open-source NGSPICE [9] based model has been proposed which will motivate researchers to develop and simulate the hybrid CMOS/MTJ circuits and systems without the need of EDA tools access. In addition, for developing hybrid circuits with random process fluctuations, a simulator-independent Monte-Carlo simulation capability has been incorporated.

This paper is divided into the following sections: The MTJ structure and working principles are explained in Section II. Section III describes the STT/SHE-MTJ physics to be modeled. Section IV contains model description and NGSPICE implementation. Section V presents the simulation and results. Section VI contains the spiking neuron MTJ design and finally, Section VII contains the conclusion, which brings the work to a close.

## II. MTJ STRUCTURES AND WORKING PRINCIPLES

The multilayer structure of the MTJ is made up of both ferromagnetic and non-magnetic materials. Fig. 1(a) illustrates a two-terminal, three-layer structure with the tunnel barrier (TB) layer sandwiched between the free layer (FL) and reference layer (RL), whereas Fig. 1(b) demonstrates a three-terminal, four-layer structure with an additional Hall material layer (HM). The reference layer, which is made up of CoFeB, offers a fixed magnetization. The tunnel barrier has MgO as dielectric, which has a typical thickness of few nanometers [7]. The free layer is the data-storage layer, whose magnetization direction can be controlled by spin-polarized current. Tunneling magnetoresistance (TMR) is a change of the tunneling current in MTJ when relative magnetization of two ferromagnetic layers changes their alignment. When the magnetization in the FL is parallel (anti-parallel) to the RL, the MTJ's resistance becomes low (High) due to TMR effect.

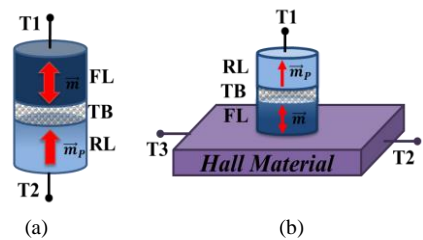

Fig. 1. Basic structure of MTJ (a) STT-MTJ (b) SHE-MTJ

The magnetization of the MTJ FL can be flipped using various methods such as voltage-controlled magnetic anisotropy (VCMA), field-induced magnetic switching (FIMS), thermally assisted switching (TAS), spin transfer torque (STT) and spin-Hall effect (SHE), where STT and SHE are the most commonly used switching techniques.

### A. Spin Transfer Torque (STT)

In the spin transfer torque (STT) switching mechanism the charge current goes through the MTJ fixed layer, polarizes, and exerts a torque on the free layer. By applying considerable torque to the free layer, it is possible to modify its magnetization by 180 degrees. In addition, as shown in Fig. 2. the bidirectional current is required for the switching of free layer magnetization. If the current magnitude exceeds a threshold current ($I_c$), STT flips the magnetization. The threshold current $I_c(AP{\rightarrow}P)$ may differ significantly from $I_c(P{\rightarrow}AP)$ [7] due to the bias dependency of STT efficiency, internal/external field and field like torque.

In addition, the average switching time ($t_w$), which is inversely proportional to the write current, is a crucial parameter. In other words, higher write current over the threshold $I_c$, results in the faster the switching of FL magnetization. Further, depending on the amplitude and duration of the write current pulse, $t_w(AP{\rightarrow}P)$ might differ from $t_w(P{\rightarrow}AP)$. Since no external current carrying conductors are required to create the external magnetic field, this approach has the advantages of scalability and high density as compared to magnetic field-based magnetization switching.

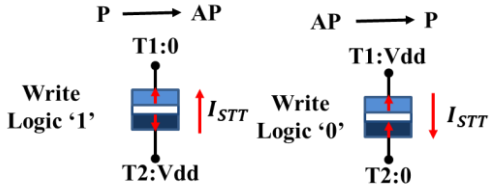

Fig. 2. MTJ switching through STT techniques

### B. Spin-Hall Effect (SHE)

A bidirectional current passing through a heavy metal strip of spin-Hall material (SHM) causes the free layer of the SHE-MTJ to switch as shown in Fig. 3. SHE can attain high efficiency of spin torque even greater than 100% when compared to the STT. Further it has a low write current that flows through the SHM. In addition to their three-terminal MTJ, the read path is separated from the write path. This separation significantly enhances the reliability of the device in terms of read operation.

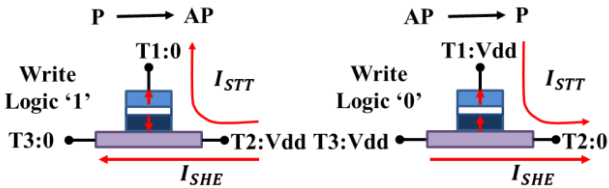

Fig. 3. MTJ switching through SHE techniques

### III. STT/SHE-MTJ PHYSICS TO BE MODELED

This section presents a SPICE compatible MTJ model that is used to emulate the behaviour of STT/SHE-MTJ structures using the Stoner-Wohlfarth monodomain approximation [10].

### A. Magnetic Anisotropy Module

Magnetic Anisotropy (MA) decides the preferential direction of magnetization in magnetic materials. It is commonly called as the energy required to move the magnetization away from its easy axis. magnetic anisotropy is classified into different types based on the alignment of their easy axis and the different sources of magnetic anisotropy. There are two kinds of magnetic anisotropy: in-plane magnetic anisotropy (IMA) and perpendicular magnetic anisotropy (PMA). In-plane anisotropy arises when the easy axis is aligned with the plane of the magnet. The in-plane magnetization aligns along the longest dimensions also the demagnetization energy/field tries to keep the magnetization in the plane of a nano-layer (free layer). The perpendicular anisotropy tries to align the magnetization perpendicular to the plane of the magnetic nano-layer as shown in Fig. 4. This type of anisotropy is caused by crystalline or interface anisotropy. The interface anisotropy comes into existence when a ferromagnetic electrode interfaces with a tunnel barrier where the thickness $t_{FL}$ of the ferromagnetic layer is less than the critical thickness ($t_c$). The perpendicular anisotropy field is also given as:

$$H_d = \frac{2K_{eff}}{M_S} \qquad (1)$$

Where, $K_{eff}$ is the effective crystal anisotropy of material, $H_d$ is the effective shape anisotropy field. Also, where $K_{eff}$ is defined as follows:

$$K_{eff} = K_\parallel + K_\perp \qquad (2)$$

where,

$$K_\parallel = 2\pi(N_{dy} - N_{dx})M_s^2 \qquad (3)$$

$$K_\perp = K_b + \frac{K_i}{t_{FL}} - 2\pi N_{dz}M_s^2 \qquad (4)$$

Where, $M_S$ is the saturation magnetization, $N_{dx}, N_{dy}, and\ N_{dz}$ are shape-dependent demagnetizing tensor components, $K_i$ for interfacial anisotropy density, and $t_{FL}$ is the free layer thickness (FL). Herein, $K_b$ bulk anisotropy density is assumed to be zero.

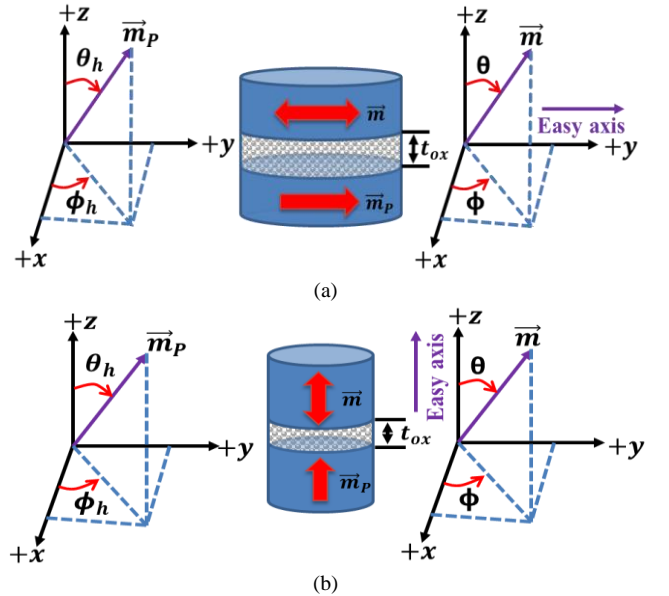

Fig. 4. Illustration of the free layer and reference layer along with magnetization components (a) IMTJ and (b) PMTJ

In an elliptical-shaped IMTJ device, a significantly high value of $N_{dz}$ is needed to retain the magnetization of the free layer in the easy plane. With a thicker free layer ($> 2$ nm) that exhibits significantly less interfacial anisotropy density ($K_i/t_{FL}$) as compared to the demagnetization factor ($2\pi N_{dz}M_s^2$), resulting in a negative value of perpendicular anisotropy ($K_\perp$). The magnetization of an MTJ device with $t_{FL} < t_C$ becomes perpendicular to the plane of the free layer. Due to the same values of $N_{dx}$ and $N_{dy}$ in the cylindrical PMTJ device, $K_\parallel$ becomes zero. However, $K_i/t_{FL}$ rises much higher than the demagnetizing factor resulting in an exceptionally high $K_\perp$ value. In a PMTJ device, the magnitude of $K_{eff}$ is around one order higher than in an IMTJ device[10].

### B. Implementation of Free Layer Magnetization Dynamics

Vector representation of magnetization of the FL ($\vec{m}$) and magnetization of the RL ($\vec{m}_P$) shown in Fig. 5 can be written as,

$$\vec{m} = \sin\theta\cos\phi\, a_x + \sin\theta\sin\phi\, a_y + \cos\theta\, a_z \tag{5}$$

$$\vec{m}_P = \sin\theta_h\cos\phi_h\, a_x + \sin\theta_h\sin\phi_h\, a_y + \cos\theta_h\, a_z \tag{6}$$

The magnetization dynamics of FL ($\vec{m}$) under the influence of effective magnetic field ($H_{eff}$) can be represented using the LLGS equation as

$$\frac{d\vec{m}}{dt} = \alpha \times \left(\vec{m} \times \frac{d\vec{m}}{dt}\right) - \gamma \cdot \vec{m} \times (\vec{m} \times H_{eff}) \tag{7}$$

where $\alpha$ and $\gamma$ are the gilbert damping constant and gyromagnetic ratio, respectively. Here $H_{eff}$ is the vector sum of crystalline anisotropy field ($H_k$), demagnetization field ($H_d$), thermal field ($H_{th}$), and spin torque field ($H_S$) can be expressed as

$$H_{eff} = H_k + H_d + H_{th} + H_S \tag{8}$$

The crystalline anisotropy field ($H_k$) coincide with the easy axis as illustrated in Fig. 4. The demagnetization field ($H_d$) can be represented as in-plane ($H_{dx}$ and $H_{dy}$) and perpendicular plane ($H_{dz}$) components

$$H_{dxy} = 4\pi(N_{dy} - N_{dx})M_S \tag{9}$$

$$H_{dz} = 4\pi N_{dz} M_S \tag{10}$$

Here, thermal noise field is presented as an external magnetic field component ($H_{th}$). The spin-torque equivalent magnetic field ($H_S$) created by current density ($J$) has two components: $H_{STT}$ and $H_{FLT}$, the magnetic fields associated with STT and FLT, respectively.

$$H_S = H_{STT} + H_{FLT} \tag{11}$$

The $H_{STT}$ can be modeled as

$$H_{STT} = \frac{\hbar \eta_{STT} J}{2qM_S V H_k}(m \times m_P) \tag{12}$$

Where, $\hbar$ is the reduced plank's constant, $q$ is the electron charge, and $V$ is the volume of an MTJ's FL. Herein, $\eta_{STT}$ is the spin polarization efficiency, which is the depends on spin polarization factor ($P$) can be expressed as

$$\eta_{STT} = \frac{P}{2(1+P^2(m \cdot m_P))} \tag{13}$$

In an IMTJ device, the magnitude of a $H_{FLT}$ can be as high as 30-40% of the STT [10], but it is insignificant in a PMTJ device. The $H_{FLT}$ can be specified as

$$H_{FLT} = \frac{\hbar \eta_{FLT} J}{2qM_S V H_k} m_P \tag{14}$$

The spin polarization efficiency $\eta_{FLT}$ can be modeled as

$$\eta_{FLT} = f_P \eta_{STT} \tag{15}$$

Where, $f_P$ denoted as the pre-factor for the contribution of FLT to the STT, which can be in range from 0.1-0.4 for an IMTJ device. In an IMTJ device, the FLT can aid the magnetization switching process. The LLGS equation can also be expressed in a generalized form as

$$\frac{1+\alpha^2}{\gamma H_k}\begin{bmatrix} \frac{d\theta}{dt} \\ \frac{d\phi}{dt} \end{bmatrix} = \boldsymbol{\tau}_u + \boldsymbol{\tau}_d + \boldsymbol{\tau}_{th} + \boldsymbol{\tau}_{STT} + \boldsymbol{\tau}_{FLT} \tag{16}$$

here $\boldsymbol{\tau}$ is the torque associated with the relevant magnetic field as defined in equations (8) and (11). For a PMTJ device, the time derivative components of respective magnetic fields can be expressed as

$$\boldsymbol{\tau}_u = \begin{bmatrix} -\alpha\sin\theta\cos\theta \\ -\cos\theta \end{bmatrix} \tag{17}$$

$$\boldsymbol{\tau}_d = -h_{ep}\begin{bmatrix} \sin\theta\cos\phi(\sin\phi + \alpha\cos\theta\cos\phi) \\ \alpha\sin\phi - \cos\theta\cos\phi \end{bmatrix} \tag{18}$$

$$\boldsymbol{\tau}_{th} = h_{th}\begin{bmatrix} (\sin\phi + \alpha\cos\theta\cos\phi)H_x \\ +(\alpha\cos\theta\sin\phi - \cos\phi)H_y - \alpha\sin\theta H_z) \\ \frac{1}{\sin\theta}(\cos\theta\cos\phi - \alpha\sin\phi)H_x \\ +\frac{1}{\sin\theta}(\cos\theta\sin\phi + \alpha\cos\phi)H_y - H_z \end{bmatrix} \tag{19}$$

$$\boldsymbol{\tau}_{STT} = h_{STT}\begin{bmatrix} (\cos\theta\cos\phi - \alpha\sin\phi)H_{sx} \\ +(\cos\theta\sin\phi + \alpha\cos\phi)H_{sy} - \sin\theta H_{sz} \\ \frac{1}{\sin\theta}(-\alpha\cos\theta\cos\phi - \sin\phi)H_{sx} \\ +\frac{1}{\sin\theta}(\cos\phi - \alpha\cos\theta\sin\phi)H_{sy} + \alpha H_{sz} \end{bmatrix} \tag{20}$$

$$\boldsymbol{\tau}_{FLT} = h_{FLT}\begin{bmatrix} (\cos\theta\cos\phi - \alpha\sin\phi)H_{sx} \\ +(\cos\theta\sin\phi + \alpha\cos\phi)H_{sy} - \sin\theta H_{sz} \\ \frac{1}{\sin\theta}(-\alpha\cos\theta\cos\phi - \sin\phi)H_{sx} \\ +\frac{1}{\sin\theta}(\cos\phi - \alpha\cos\theta\sin\phi)H_{sy} + \alpha H_{sz} \end{bmatrix} \tag{21}$$

where,

$h_{ep} = \frac{4\pi M_S}{H_k}$, $h_{th} = \frac{H_{th}}{H_k}$, $h_{STT} = \frac{\hbar \eta_{STT} J}{2qM_S V H_k}$, $h_{FLT} = \frac{\hbar \eta_{FLT} J}{2qM_S V H_k}$

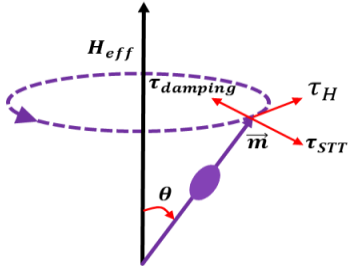

Fig. 5. Free layer magnetization components

## C. Implementation of TMR and MTJ Resistance

There is a magnetization-dependent resistance-associated between the two terminals of the MTJ device. This magnetization-dependent resistance is modeled and incorporated inside the MTJ model as a SPICE subcircuit. Herein, the TMR is determined by the spin polarization ($P$) and the bias voltage across the MTJ ($V_{mtj}$), which is represented as

$$TMR(V,T) = \frac{2P(T)^2}{1-P(T)^2}\left\{\frac{1}{1+\left(V_{mtj}/V_0\right)^2}\right\} \quad (22)$$

wherein $P(T) = P_0(1 - \alpha_{sp}T^{3/2})$. The effect of saturation magnetization is also taken into account $M_s(T) = M_{s0}(1 - T/T_c)^\beta$. Here $P_0$ and $M_{s0}$ represents the spin polarization and saturation magnetization at 0K, $T$ represents absolute temperature. $T_C$ represents the Curie's temperature, $\alpha_{sp}$ and $\beta$ represents the material dependent constant. The tunnel magnetoresistance ratio (TMR) at zero bias is defined as $TMR_0 = (R_{AP} - R_P)/R_P$. The resistance-area product (RA) of the device determines the parallel resistance $R_P$ of an MTJ. The parallel and antiparallel resistance of an MTJ, i.e., $R_P$ and $R_{AP}$ are determined using the following relationship

$$R_P = RA/(I_x * I_y) \quad (23)$$

$$R_{AP} = (1 + TMR)R_P \quad (24)$$

$$R_{MTJ} = \frac{(1+\cos\theta)(R_P-(TMR+1)R_P)}{2} + (TMR+1)R_P \quad (25)$$

where $V_0$ represents the fitting parameter. Once the $RA$ value is determined, $R_P$ and $R_{AP}$ are computed by considering MTJ area and TMR in the proposed model. In the case of SHE-MTJ modeling, the effective resistance is calculated by the MTJ resistance as well as the spin-Hall material resistance, as illustrated in Fig. 6.

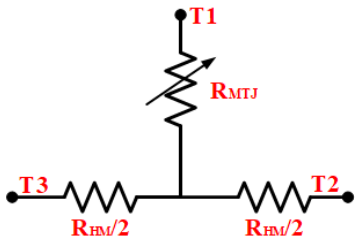

Fig. 6. SHE resistance modeling

Hall material resistance $R_{HM}$ is expressed as,

$$R_{HM} = \rho_{HM}\frac{l}{w.d}, \quad R_{SHE-MTJ} = R_{MTJ} + \frac{R_{HM}}{2} \quad (26)$$

where $\rho_{HM}$ is the Hall material resistivity and $l$, $w$, $d$ are the dimensions of Hall material. $R_{SHE-MTJ}$ is the effective resistance of the SHE-MTJ.

## D. NGSPICE Noise Implementation

The stochastic thermal noise which depends on the temperature can affect the magnetization dynamics on the MTJ. Using a random field, stochasticity of magnetization switching is included in the proposed model, which helps to accurately model the thermal noise effect on MTJ switching. Thermal noise has been implemented in this model using a resistor in parallel with a thermal noise current source, resulting in the computation of random thermal field $H_{th}(x,y,z)$ [11].

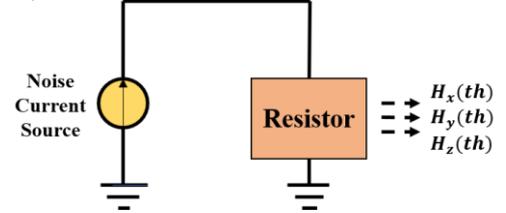

Fig. 7. Thermal noise equivalent circuit diagram

This random thermal field is added into the LLGS equation, which implements the stochasticity thermal fluctuations induced in MTJ, resulting in a stochastic LLGS equation. The magnitude of the random thermal field is given by:

$$H_{th} = \sqrt{\frac{\alpha}{1+(\alpha^2)}\frac{2k_BT}{\gamma\mu_0 M_s V\delta(t)}} \quad (27)$$

Where, $\delta(t)$ represents the simulation time step, $k_B$ represents the Boltzmann's Constant, $\alpha$ represents the damping ratio, and $\gamma$ represents the gyromagnetic ratio.

The influence of stochastic thermal variations on MTJ switching is taken into consideration by the SPICE model, similar approach as used in earlier model [10]. As shown in Fig. 9, the noise is implemented in NGSPICE using three TRNOISE current sources that have a gaussian distribution with a zero mean and unit standard deviation.

## E. Modeling the Spin-Hall Effect

SHE converts a charge current that is unpolarized into a pure or 100% polarized spin current [12]. This happens when the charge current travels through the spin-Hall material having spin-orbit coupling interactions [13]. The SHE can produce enough torque to alter the magnetization of the MTJ free layer, which is present on top of the SHM layer. As demonstrated in Fig. 8, the switching orientation of the FL is determined by a bidirectional current in SHM.

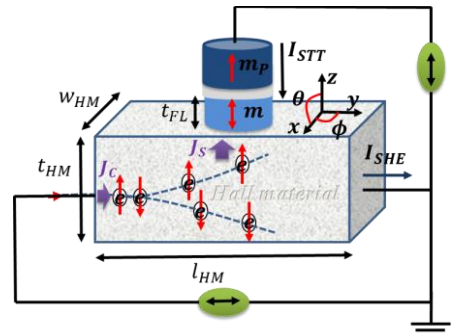

Fig. 8. Spin-polarized current in SHE-MTJ

$P_{SHE}$ represents spin polarization ratio which is dependent on the geometry of MTJ and SHM as well as the bulk spin-Hall angle $\theta_{SHE}$.

$$P_{SHE} = \frac{I_{spin}}{I_{charge}} = \frac{A_{MTJ}}{A_{SHE}} \times \frac{J_{MTJ}}{J_{SHE}} \quad (28)$$

$$P_{SHE} = \frac{A_{MTJ}}{A_{SHE}} \times \theta_{SHE0}\left(1 - \frac{1}{\cosh\left(\frac{t_{SHM}}{\lambda_{SHM}}\right)}\right) \quad (29)$$

$$\theta_{SHE} = \frac{J_{MTJ}}{J_{SHE}} = \theta_{SHE0}\left(1 - \frac{1}{\cosh\left(\frac{t_{SHM}}{\lambda_{SHM}}\right)}\right) \quad (30)$$

Here, the cross-sections and current density of the MTJ(SHM) are represented by $A_{MTJ}$ ($A_{SHE}$) and $J_{MTJ}$ ($J_{SHE}$) respectively. $\theta_{SHE0}$ is the bulk spin-Hall angle, whereas $t_{SHM}$ and $\lambda_{SHM}$ are the thickness and spin diffusion length of SHM, respectively.

## IV. MODEL DESCRIPTION AND NGSPICE IMPLEMENTATION

This section contains a description of the model as well as its NGSPICE implementation. As illustrated in Fig. 9, the proposed model incorporates different subcircuits that are integrated to replicate complete STT/SHE MTJ behavior.

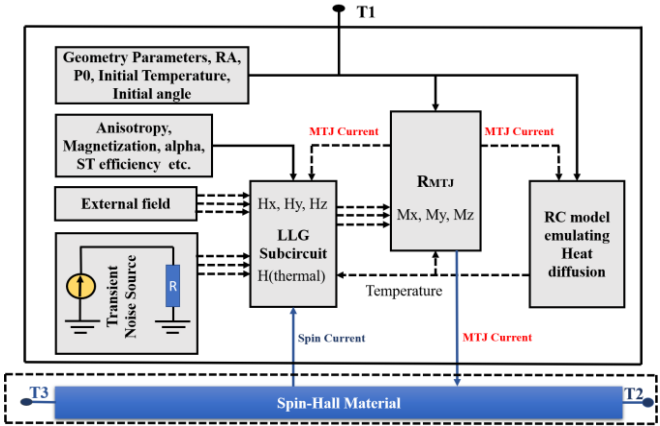

Fig. 9. NGSPICE simulation framework for STT-MTJ and SHE-MTJ Models

### A. STT-MTJ Model Description

The SPICE subcircuits used for NGSPICE MTJ model are discussed in this section. The proposed model uses dedicated subcircuits to emulate different aspects of device behavior such as LLG subcircuit for magnetization switching, $R_{MTJ}$ subcircuit for computing the resistance of MTJ which is a function of magnetization, temperature, and bias voltage, whereas heat diffusion subcircuit for emulating the influence of self heating in MTJ devices [19]. To make it compatible with the NGSPICE simulator, more dependent sources have been incorporated. LLG subcircuit implements the LLGS equation with both spin-torque, external magnetic field effects, and random thermal field. Similar to [14], [20]. The torque exerted by each component of the LLGS equation is equivalent to a pair of capacitors charged by voltage-dependent current sources. Herein, voltage is equivalent of magnetization vector in polar coordinates.

### B. SHE-MTJ Model Description

The simulation framework for the SHE-MTJ model comprises of four subcircuits in which an additional fourth subcircuit represents the behavior of spin-Hall material as shown in Fig. 9. Here, the modified LLGS solver and magnetization dependent resistance as used in the STT-MTJ model comprises the spin-Hall effect. To integrate the system's temperature and resistance with the spin's dynamic motion, the subcircuits are merged into a single MTJ subcircuit model file. Based on the block diagram in Fig. 9, explanations and the basic physics behind each block model are discussed in section III.

TABLE I

| List of User-Defined Parameter | |
| --- | --- |
| **Description** | **Value** |
| FL dimensions, $l_x \times l_y \times l_z$ | **65nm×65nm×1.48nm** |
| SHM dimensions, $l_{xshm} \times l_{yshm} \times l_{zshm}$ | **80nm×65nm×2.2nm** |
| Saturation Magnetization, $M_{Os}$ | 1300 emu/cm$^3$ |
| Anisotropy Constant $K_u$ | 651600 erg/cm$^3$ |
| Polarization Factor, $P_0$ | 0.85 |
| Damping Factor, $\alpha$ | 0.01 |
| Temperature, $TMP_0$ | 300K |
| Simulation Time Step, $\delta(t)$ | 1ps |
| Uniaxial field strength, $H_K$ | 100 Oe |
| Resistance area product, $RA$ | 5.4 Ω μm$^2$ |
| **List of Model Constant** | |
| Gyromagnetic Ratio, $\gamma$ | $1.76 \times 10^7$ Oe$^{-1}$s$^{-1}$ |
| Boltzmann Constant, $K_B$ | $1.38 \times 10^{-16}$ erg/K |
| Elementary Charge, $q$ | $1.6 \times 10^{-19}$C |
| Hall material resistivity | 200 μΩ. cm$^2$ |
| Spin-Hall Angle, $\theta_{she}$ | 0.3 |

## V. SIMULATION AND RESULTS

To demonstrate and analyze the performance of the proposed MTJ models, NGSPICE circuit simulations were performed. Here, PMTJ with dimensions of 65 nm × 65 nm × 1.48 nm is chosen. Table I shows the other parameters and constants used in this model for both STT/SHE-MTJ models. Various analysis is performed to compute the performance and characteristic of the MTJ. For this, the variation of FL magnetization of dynamics, stochastic behaviour, MTJ resistance with bias voltage, and thermal noise effect is shown.

### A. STT/SHE-MTJ FL Layer Spin Dynamics

To show the usefulness of the model in designing hybrid MTJ/CMOS and other circuits, NGSPICE is used to perform transient analysis on a single STT/SHE-PMTJ device. The simulation results are shown in Fig. 10 for STT and Fig. 11 for the SHE switching mechanism, wherein MTJ is in the AP state initially with zero bias is applied. Then, to demonstrate AP(P) to P(AP) switching, a positive(negative) bias voltage is applied. TMR dependence on applied bias voltage is also seen in Fig. 10.

### B. MTJ Resistance

To show the bias voltage dependency and temperature effect on MTJ resistance, transient analyses are performed at different operating temperatures with applied voltage. The resistance area production ($RA$) decided the parallel resistance ($R_P$) here it is 5.4. The simulation result is shown in Fig. 12 demonstrating the effect of bias voltage and the temperature on MTJ resistance.

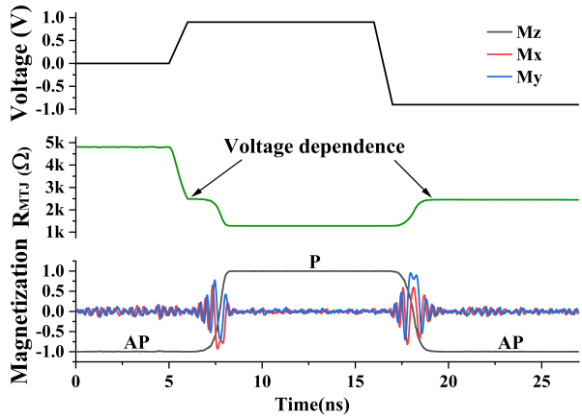

Fig. 10. Timing diagrams demonstrate the magnetization of the FL, biasing voltage, and STT-MTJ's resistance.

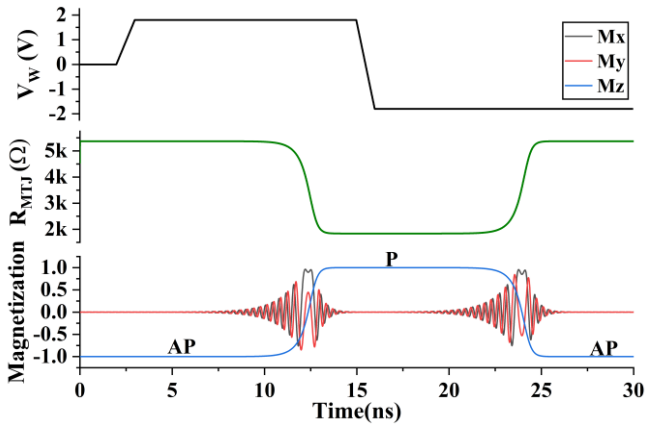

Fig. 11. Timing diagrams demonstrate the magnetization of the FL, biasing voltage, and SHE-MTJ's resistance.

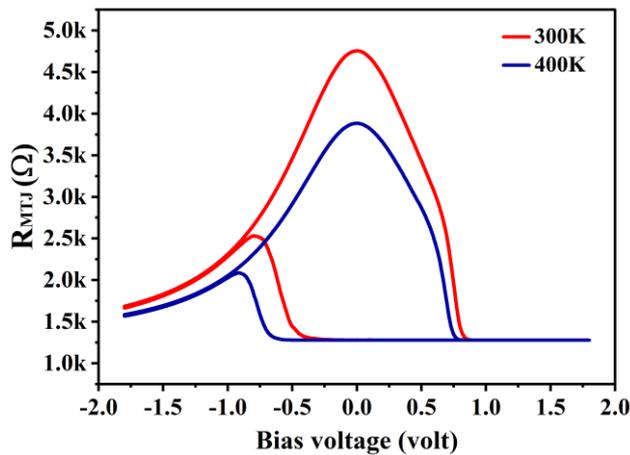

Fig. 12. MTJ resistance variation with temperature and bias voltage using the proposed framework

### C. Stochastic Switching Behavior of MTJ

To illustrate the stochastic nature of the MTJ switching, four successive simulations have been carried out initialized using a different random seed as shown in Fig. 13, At 300K, the switching duration changes owing to thermal variations. MZ denotes the magnetization of the free layer along its z-axis.

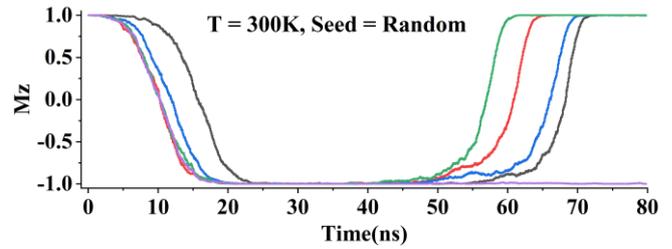

Fig. 13. The timing diagram demonstrates the MTJ's stochastic switching behaviour at 300K

### D. Process Variations / Monte-Carlo

Random process variations that occur during the manufacturing of devices like MTJs can result in a significant change in circuit performance. Predicting the impact of these random variations early in the design process can make design flow much more efficient in terms of cost and time to commercial product. It is very difficult to analytically predict the random variations that can occur while manufacturing therefore one can use Monte-Carlo models on the geometry variations of the device. In this paper, Monte-Carlo simulation for STT MTJ read and write operation is carried out on the circuits shown in Fig. 14 and Fig. 16, respectively. The impact of random variations is demonstrated for the read operation on pre-charged sense amplifier (PCSA) are shown in Fig. 15. Here MTJ's physical parameters such as length, width, height of free layer, initial polarization ($P_0$), etc. can take random values from a gaussian distribution to emulate random process variations. Monte-Carlo results shows ~40 failures for PCSA read in 1000 runs. Further, the average read current in P and AP state of MTJ is 53uA and 39uA respectively. Also, for write operation (on circuit in Fig. 16) the impact of random variations are demonstrated in Fig. 17. Its Monte-Carlo results shows ~92 failures in the case of STT MTJ. Similar analysis carried out on SHE MTJ resulted in ~18 failures for PCSA read and ~72 failures for write operation.

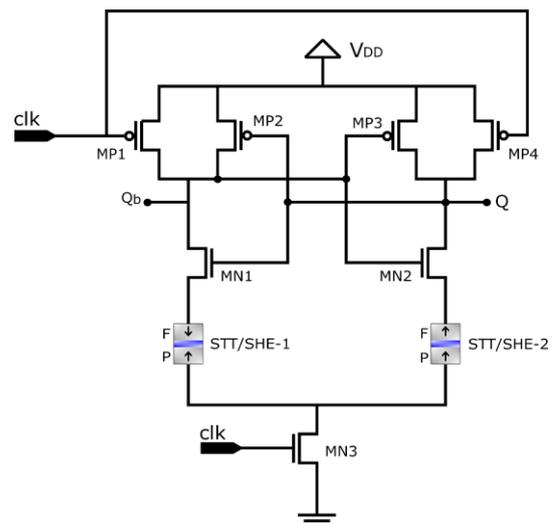

Fig. 14. Schematic diagram of PCSA read using STT/SHE

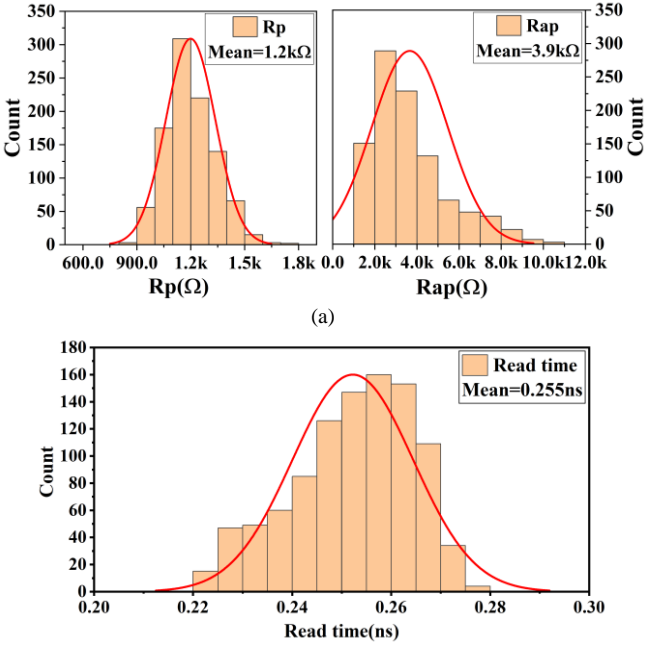

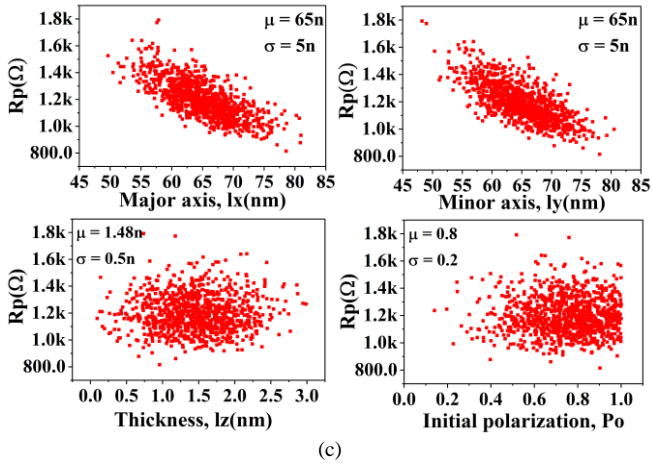

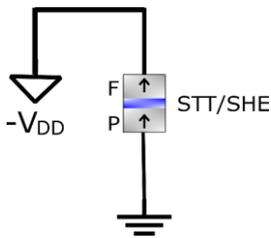

Fig. 15. Monte-Carlo simulation for STT-MTJ based PCSA read operation (a) Parallel ($R_P$) and anti-parallel ($R_{AP}$) resistance (b) Read time (c) Parallel resistance variation plots w.r.t free layer dimensions and initial polarization

Fig. 16. MTJ writes from P to AP configuration through STT/SHE

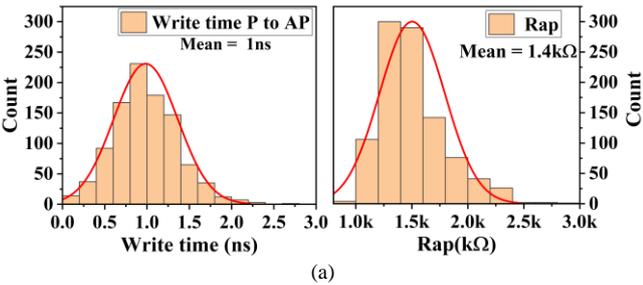

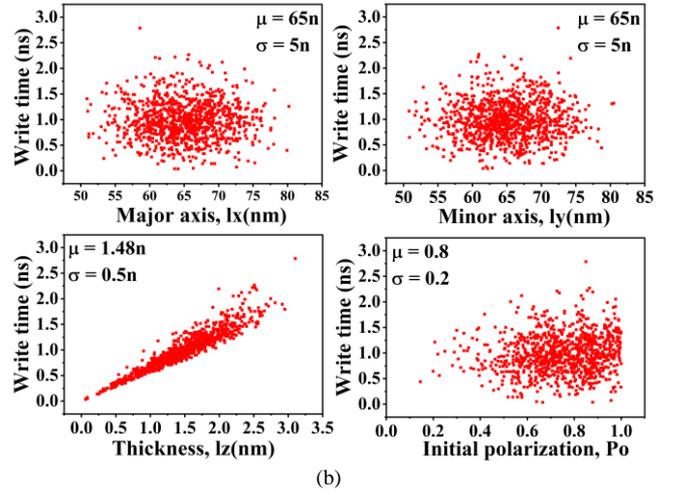

Fig. 17. (a) Statistical distribution of write time (ns) and $R_{AP}$ (kΩ) (b) P→AP switching delay (ns) in STT-MTJ with diferent free layer dimension (nm) and initial polarization ($P_0$).

### E. SHE assisted STT-MTJ Switching Analysis

The switching time of the FL magnetization has been demonstrated when SHE assisted STT switching in Fig. 18 which shows the reduction in switching time with to 1ns. With SHE, it has been around a 50 % reduction in delay, which leads to a huge advantage when it comes to memory applications.

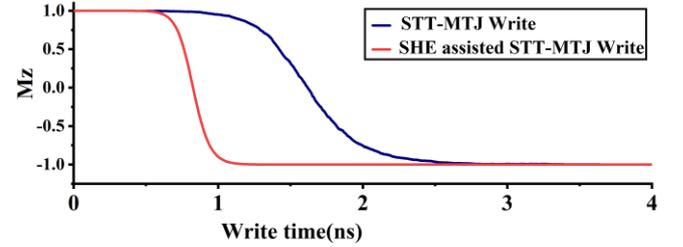

Fig. 18. Magnetization switching of MTJ with the STT and SHE mechanism.

### F. Comparison with Previously Proposed Models

In this subsection a comparative study with other MTJ models is discussed. The NGSPICE based STT/SHE-MTJ model has the following benefits over other standard models, as presented in Table II.

TABLE II
COMPARISON WITH THE PREVIOUSLY PROPOSED MODELS

| MTJ model Tools | Type | Monte Carlo Sim. Capability | Open-source | STT and SHE compatibility |
|---|---|---|---|---|
| Synopsis Sentaurus device | C/C++ based [8] | dependent | No | No |
| Hspice | SPICE Subcircuit model [13],[14],[16] | dependent | No | Yes |
|  | Verilog-A [7], [10] | dependent | No | Yes |
| Cadence | Verilog-A [6], [15], [17] | dependent | No | No |
| NGSPICE | SPICE Subcircuit model | Independent | Yes | Yes |

## VI. SPIKING NEURON MTJ DESIGN

To demonstrate the capability of integrating the SHE and STT-MTJ compact models with CMOS for circuit-level simulation, a spiking stochastic neuron based on MTJ as shown in [5] is implemented. The circuit is shown in Fig. 19(a), in which a charge current travels through HM with high-spin orbit coupling produces spin accumulation at the HM-ferromagnetic free layer interface (FL). The three-terminal SHE-MTJ devices function as stochastic neurons with independent current paths for "read" and "write." The transistors ($T_W$) divide the current channels for "write" and "read." During the "write" cycle (Transistor $T_W$ is ON), the incoming charge current, $I_{SYN}$, switches the neuron probabilistically based on its amplitude in the presence of thermal noise. A small read current $I_{READ}$ runs through the "Reference" and "Neuron" MTJs in succession during the succeeding "read" cycle (Transistor $T_R$ is ON). Initially, the magnetization of "Reference" MTJ's is set to the *AP* state, allowing the inverter to emit a spike ($V_{SPIKE}$) if the "Neuron" MTJ transitions from the *P* to the *AP* state. The "Neuron" MTJ is reset to the *P* state using a reset current $I_{RESET}$ if the neuron spiked. The waveform demonstrating the operation is shown in Fig. 19(b).

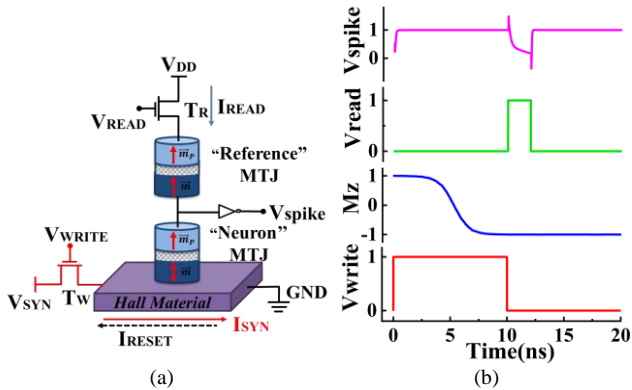

Fig. 19. (a) Neuron MTJ (b) Transient simulation of spiking neuron MTJ

## VII. CONCLUSION

This paper presents, novel STT/SHE-MTJ compact model for hybrid MTJ/CMOS circuit simulation, that accurately emulates the device physics and stochastic thermal noise behavior of the MTJ. Furthermore, the proposed model is superior from the previous models in terms of the compatibility with the open-source NGSPICE simulator and simulator-independent Monte-Carlo simulation capability. The results demonstrate the importance of the proposed model.

## VIII. ACKNOWLEDGMENT

The Science and Engineering Board (SERB) start-up grant and seed grant of IIT BHU have supported this project. The corresponding project file number for SERB is SRG/2019/000573. The authors also acknowledge the Department of Science and Technology, Government of India.